%% file: paper_eprint.tex
\documentclass[twocolumn,showpacs,aps,prl,superscriptaddress]{revtex4}

\usepackage{graphicx}
\usepackage{dcolumn}
\usepackage{amsmath}
\usepackage{epsfig}
\usepackage{array}

\input pubboard/babarsym

\input symbols

\newcommand{\BABARPubYear}    {03}
\newcommand{\BABARPubNumber}  {027}

\newcommand{\SLACPubNumber} {10141}

\def\figurebox#1#2#3{
    \def\arg{#3}
    \ifx\arg\empty
    {\hfill\vbox{\hsize#2\hrule\hbox to #2{\vrule\hfill\vbox to #1{\hsize#2\vfill}\vrule}\hrule}\hfill}
    \else
    {\hfill\epsfbox{#3}\hfill}
    \fi}

\begin{document}

\preprint{\babar-PUB-\BABARPubYear/\BABARPubNumber} 
\preprint{SLAC-PUB-\SLACPubNumber} 

\begin{flushleft}
\babar-PUB-\BABARPubYear/\BABARPubNumber\\
SLAC-PUB-\SLACPubNumber\\
\end{flushleft}

\title{
{\large \bf \boldmath 
Measurements of the Branching Fractions of Charged \B\ Decays\\ to \Kpp\ Final States} 
}

\input pubboard/authors_jul2003

\date{\today}

\begin{abstract}
We present results of searches for $B$-meson decays to 
$K^+\pi^-\pi^+$ with the \babar\ detector.
With a data sample of \bbpairs\ \BB\ pairs, 
we measure the branching fractions and 90\% confidence-level upper limits 
averaged over charge-conjugate states (the first error is statistical and the second is systematic):
$\calB(\BpmKstarpi) =  (15.5 \pm 1.8 ^{+1.5}_{-4.0})\times 10^{-6}$,
$\calB(\BpmfzK, f_0\ra \pi^+\pi^-) = \BrfzKVal$, 
$\calB(\BpmDzpi, \Dzb \ra K^+\pi^-) = \BrDzpiVal$,
$\calB(\BpmRhoK) < 6.2\times 10^{-6}$ and
$\calB$($B^+ \to K^+ \pi^- \pi^+$ non-resonant) $< 17\times 10^{-6}$.
\end{abstract}


\maketitle

The study of charmless hadronic \B\ decays can make important
contributions to our understanding of hadronic decays and \CP\ 
violation in the Standard Model~\cite{BBNS}.  Branching fraction
 predictions for $B$ meson decays to Pseudoscalar--Vector final states
 have recently been calculated using QCD Factorisation 
 and SU(3) flavor symmetry models~\cite{BN,noel,GR}. The measurement of $B^+$
 meson~\cite{conjugates} decays to the final state $K^+\pi^-\pi^+$ via
 intermediate resonances can be used to search for weak phases and
 direct \CP\ violation. The signal can be used to examine the
 light-meson mass spectrum~\cite{PDG,Minkowski:2003ja}.  The charm
 state $\chi_{c0}$ might be sensitive to the angle $\gamma$ of the
 Unitarity Triangle through interference with the non-resonant
 component producing an observable charge asymmetry~\cite{gamma2}; the
 branching fraction also constrains some models for charmonium hybrid
 production~\cite{close}.

The data used in this analysis were collected at the \pep2\ asymmetric
\epem\ storage ring with the \babar\ detector~\cite{babar}.  
The \babar\ detector consists of a five-layer silicon tracker, a drift chamber, a
new type of Cherenkov detector~\cite{DIRC}, an electromagnetic
calorimeter and a magnet with instrumented flux return. 
The data sample
has an integrated luminosity of \onreslumi\ collected at the $\FourS$
resonance, which corresponds to $(61.6\pm 0.7)\times 10^6$ \BB\ pairs
(\nbb).  We assume that the $\FourS$ decays equally to neutral and
charged $B$ meson pairs.

The \ppK\ phase space can be represented in a Dalitz plot,
in which the many resonant $B$ decay modes form overlapping bands and 
interference occurs where the bands overlap~\cite{Dalitz}.
As a consequence, the whole Dalitz plot should
be considered before assigning a branching fraction to a specific mode.
However, the data sample is not large enough for a full Dalitz plot fit 
to be effective.
In this analysis, the Dalitz plot is divided into eight regions, 
reflecting as much as possible the known decay modes.
We first determine the yields in these regions using a maximum-likelihood fit. 
We then interpret the yields as branching fractions, assuming a 
particular collection of quasi-two-body decay modes.  
We evaluate the systematic uncertainty due to the particular choice 
of decay modes and the effect of 
interference between the different contributions.

\begin{figure}[htb]
\resizebox{0.9\columnwidth}{!}{
\includegraphics{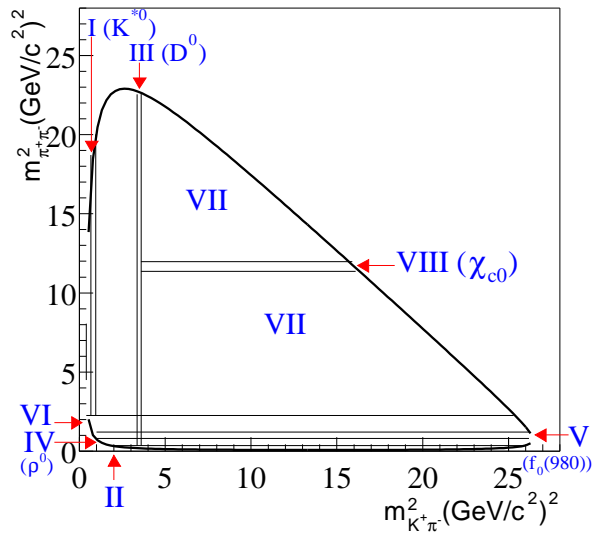}
}
\caption{A Dalitz plot showing the defined regions.
}\label{fig:Regions}
\end{figure}

The regions are defined in $m_{K\pi}$ and $m_{\pi\pi}$, the invariant masses 
of the neutral $K\pi$ and $\pi\pi$ systems, as given in 
Table \ref{tab:regions} and illustrated in Fig.~\ref{fig:Regions}.  
Region I is expected to be dominated by $K^{*0}(892) \pi^+$.  
Region II could have contributions from several higher $K^{*}$ resonances.
Region III is dominated by the production of $\Dzb \pi^+$. 
The high branching fraction for this mode allows it to be used to 
correct for differences between data and
simulated events and to evaluate systematic uncertainties.
Regions IV and V are expected to be dominated by $\rho^0(770) K^+$ 
and $\fz K^+$, respectively. 
The resonance contributions to Region VI are not known a priori.
The area where these regions intersect the \Dzb\ band, 
$1.8<m_{K\pi}<1.9 \gevcc$, is excluded.
Region VII could contain higher mass charmless and charmonium
resonances, as well as a non-resonant contribution that extends across
the whole Dalitz plot.
Region VIII is dominated by \ChiczK. This channel is vetoed from other regions
using $3.355<m_{\pi\pi}<3.475 \gevcc$.

\input dalitzTable

Candidate $B$ mesons are formed by combining three charged tracks, where
each track is required to have at least 12 hits in the drift chamber, to
have transverse momentum of at least 100~\mevc and to be consistent with
originating from the beam-spot.  Charged pions and kaons are identified
using energy loss (\dedx) measured in the tracking system and the
Cherenkov angle and number of photons measured by the Cherenkov
detector.  Pions are required to fail the kaon selection.
The efficiency of selecting kaons is approximately 80\%, 
while the probability of misidentifying pions as kaons is
below 5\%, up to a laboratory momentum of 4.0~\gevc. 
Pions are also required to fail an electron selector which uses  
\dedx, the energy to momentum ratio and the shape of the calorimeter signal.
Over 99\% of pions from the signal decay pass this requirement. 

Signal decays are identified using two kinematic variables: $\DeltaE$,
the difference between the center-of-mass (CM) energy of the $B$
candidate and $\sqrt{s}/2$, where $\sqrt{s}$ is the total CM energy;
and the beam-energy-substituted mass $\mes = \sqrt{(s/2 + \pvec_i
\cdot \pvec_B)^2/E_i^2 - \pvec^2_B}$, where $\pvec_B$ is the momentum
of the reconstructed $B$ candidate and ($E_i, \pvec_i$) is the
four-momentum of the initial $e^+e^-$ system.  The $\Delta E$ and $\mes$
distributions for signal events have widths of $20~\mev$ and
$2.7~\mevcc$, respectively.  We require $5.22<\mes<5.29 \gevcc$ and
$|\Delta E| <0.1 \gev$ for events entering the fit.  Events with $0.1<
|\Delta E| <0.3 \gev$ are used for continuum background
characterisation as described below.

A very small proportion of events, fewer than 4\%,
have two or more candidates that pass the above requirements.
For these events a single candidate is selected at random, so as not to bias the fit distributions.
This random selection has a minimal impact on the efficiency,
and any systematic uncertainty due to this effect is negligible.
A single candidate per event is similarly selected for the data with $0.1< |\Delta E| <0.3 \gev$
used in continuum background characterisation.

Continuum light-quark and charm production is the dominant source of background.
This is suppressed using two event-shape variables. 
The first is the cosine of the angle $\theta_T$ between the thrust axis 
of the selected $B$ candidate and the thrust axis of the rest of the event. 
For continuum background, the distribution of $|\cos\theta_T|$ is strongly 
peaked towards unity whereas the distribution is uniform for signal events.
We require $|\cos\theta_T| < 0.9$.
The second event-shape variable is a Fisher discriminant ($\mathcal{F}$)~\cite{Fisher}. 
For $\mathcal{F}$ we use a linear combination of the cosine of the 
angle between the $B$-candidate
momentum and the beam axis, the cosine of the angle between the $B$-candidate
thrust axis and the beam axis, and the energy flow of the rest of
the event into each of nine contiguous, concentric, $10^{\circ}$ cones around the
thrust axis of the reconstructed $B$~\cite{CLEOCones}.

There are also $B$-decay backgrounds,  mainly four-body decays, 
and three-body decays with one or more particles misidentified.  
These backgrounds are studied using Monte Carlo simulations (MC).  
They are reduced by the particle identification selections and 
by excluding events containing $\jpsi$ or $\psitwos$ decays to 
$l^+l^-$ with vetoes $2.97<m_{\pi\pi}<3.17 \gevcc$
and $3.56<m_{\pi\pi}<3.76 \gevcc$, however some $B$-decay backgrounds 
remain and must be accounted for.  
For backgrounds contributing only a few events to the maximum-likelihood (ML)
fit, the estimated contribution is subtracted from the final signal 
yield with a systematic uncertainty to account for the unknown 
probability of the background to be selected as signal in the fit. Larger 
backgrounds are parameterized in the ML fit as described below.
These were $\BpmDzpi, \Dzb \ra \Kp\pim\piz$ in regions II and
VII, $B^+ \rightarrow \Dzb \rho^+(770)$ with $\Dzb \rightarrow K^+
\pi^-$ and $\rho^+ \rightarrow \pi^+ \pi^0$ in region VII and
 $B^+ \rightarrow \eta' K^+ $ with $\eta' \rightarrow \rho^0
(770) \gamma$, $ \rho^0 \to \pi^+ \pi^-$ in regions IV and V.

We form probability density functions (PDFs) with parameters $\vec{\alpha}$ for
the three variables ($\vec{x}$) \mes, \DE, and $\mathcal{F}$ in each region.
We find the correlations among these variables to be negligible; 
accordingly, for each hypothesis $l$ (signal, continuum background, and $B$ background), 
we form a product $P_{l} = P_{l,\mes} P_{l,\DE} P_{l,\mathcal{F}}$ 
that models that hypothesis.
The likelihood for an event $j$ is the sum over the $M$ hypotheses of the products $P_{l}$,
with each product weighted by the number of events (to be determined), $n_l$, 
for that hypothesis.
A product over the $N$ events in the data sample of the event likelihoods
along with a Poisson factor forms the likelihood function:
\begin{equation}
  \label{eq:Likelihood}
  \mbox{$\mathcal{L}$} \,=\, \exp\left(-\sum_{i=1}^{M} n_i\right)\, \prod_{j=1}^N 
\,\left(\sum_{l=1}^M n_{l} \, P_{l}(\vec{\alpha},\vec{x_j})\right).
\end{equation}
 
\begin{figure}[htb]
\resizebox{\columnwidth}{!}{
\begin{tabular}{cc}
    \includegraphics{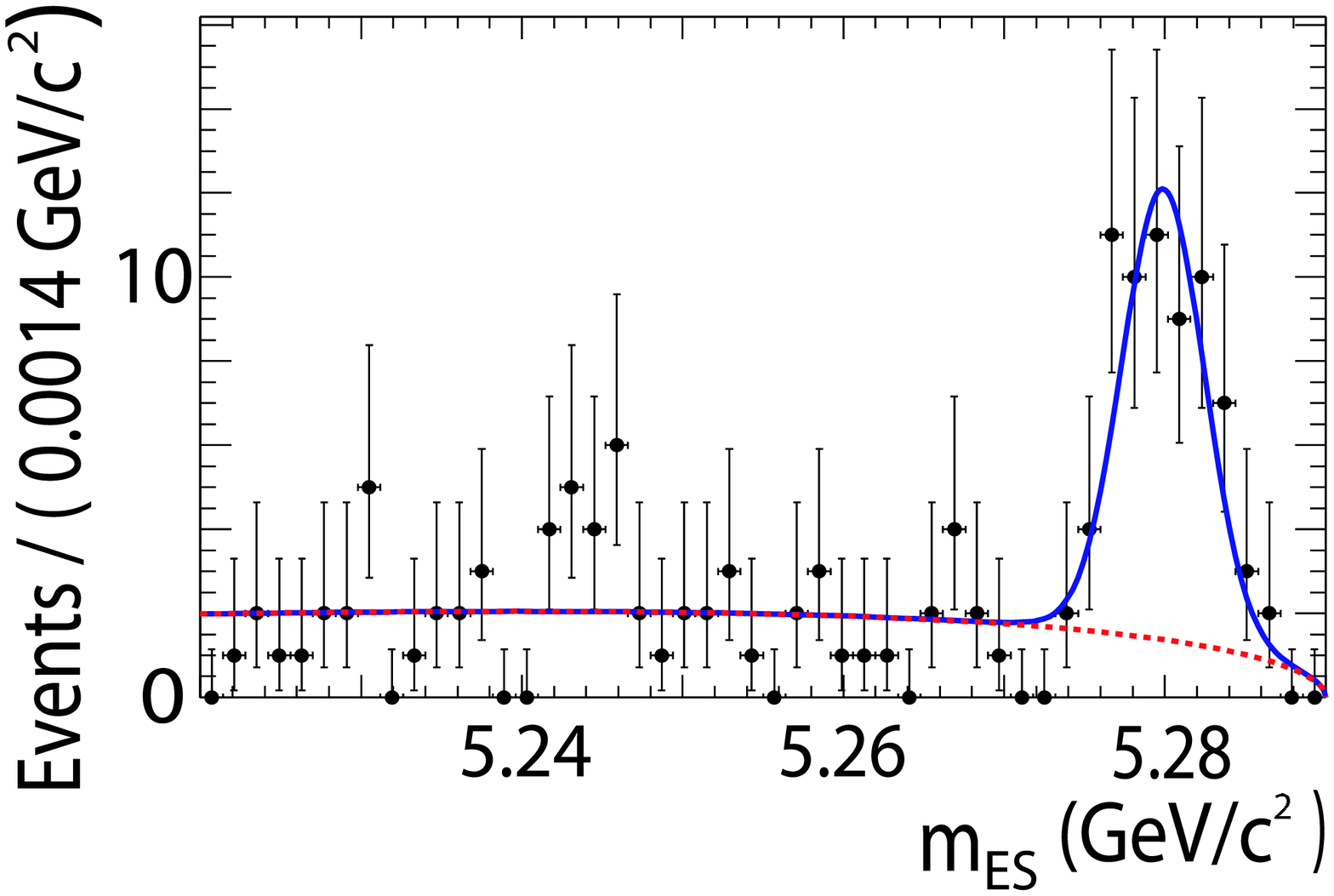} &
    \includegraphics{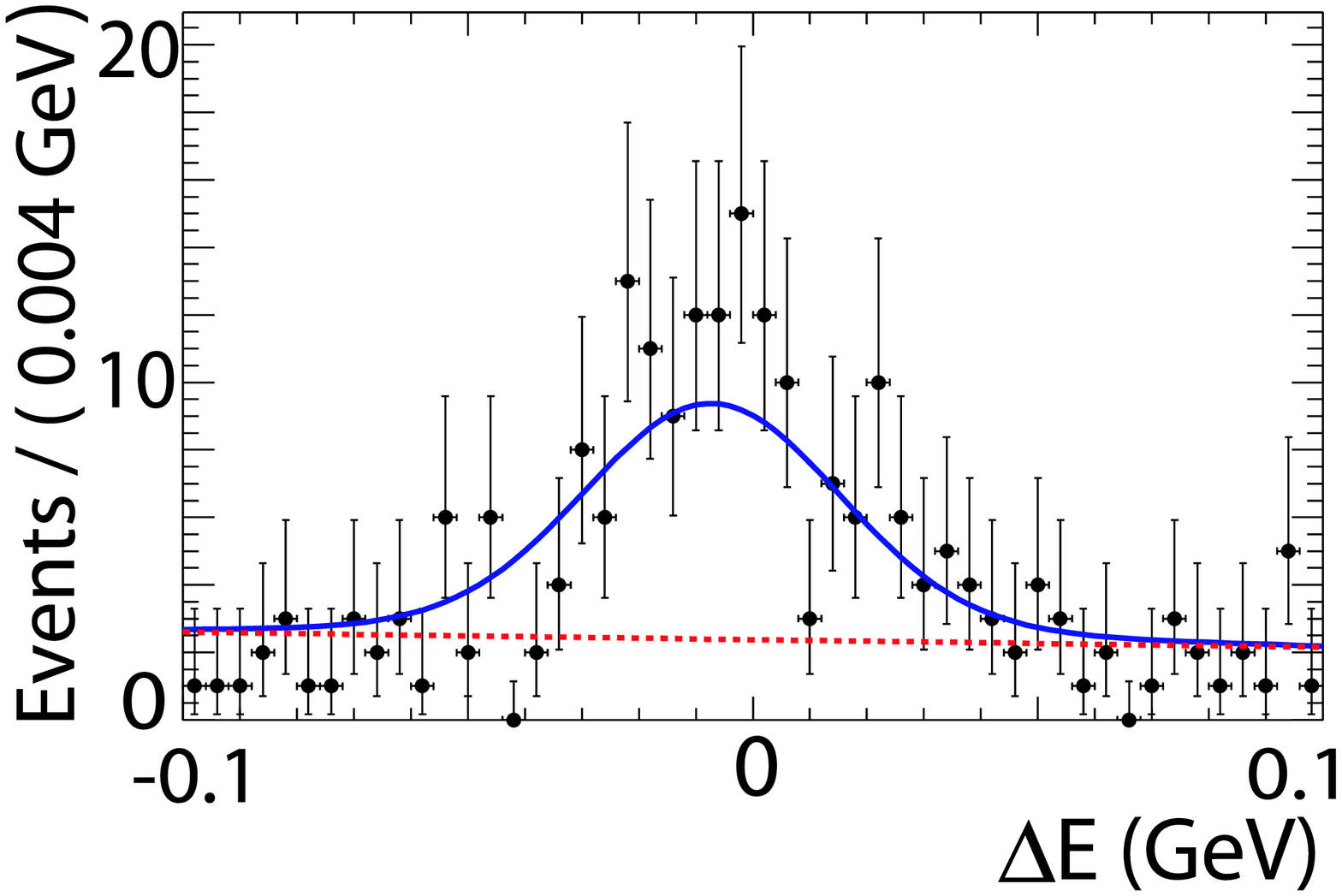} \\
    \includegraphics{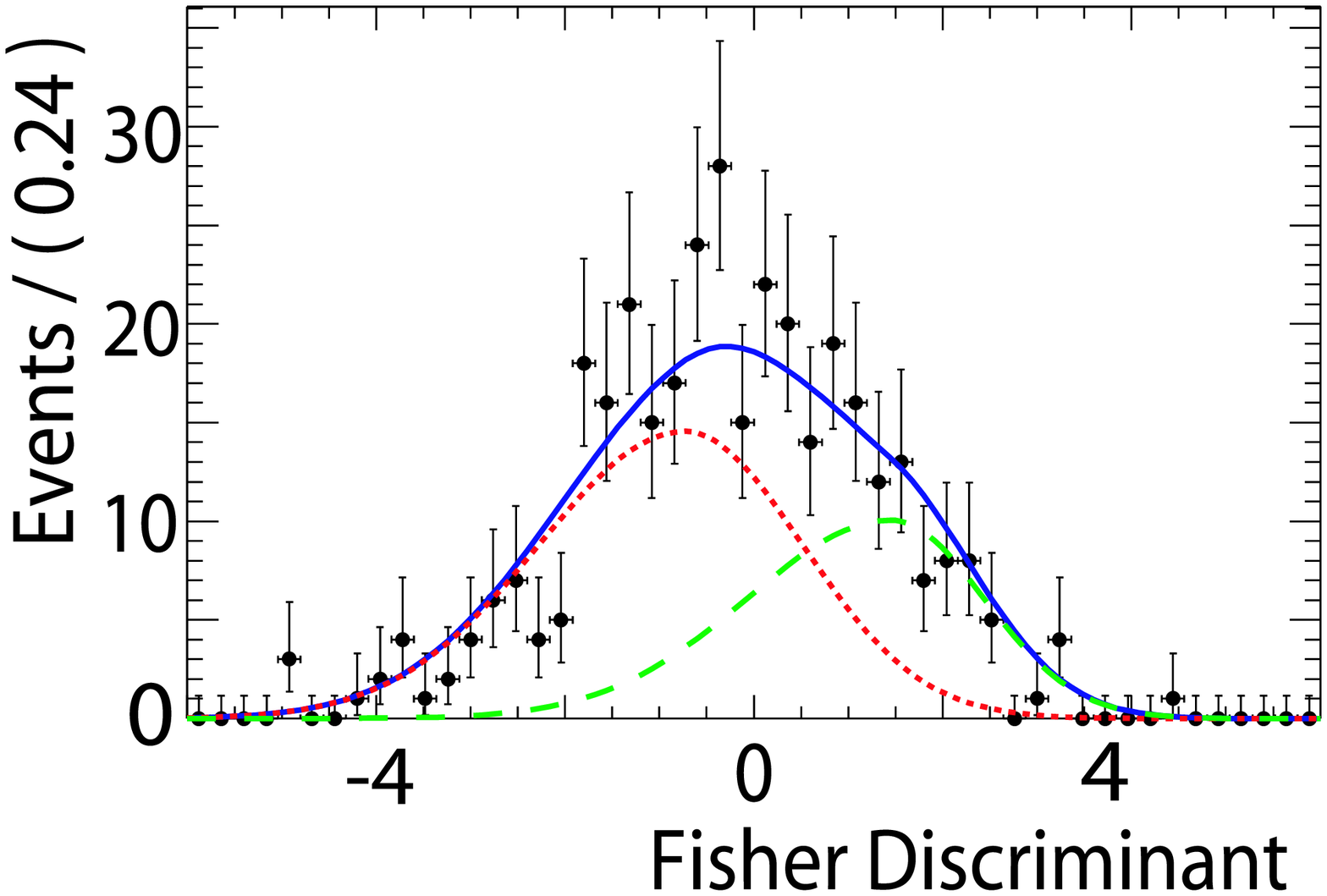} & \\
\end{tabular}
}
\caption{Projection plots in $m_{ES}$, $\Delta E$ and $\mathcal{F}$, 
for the data in Region I.
The superimposed curve is a projection of the full fit with the 
background component shown as a dotted line and, 
for $\mathcal{F}$, the signal component shown as a dashed line.
}\label{fig:mes_dE_fish}
\end{figure}

This likelihood is maximized to obtain $n_l$ for the signal and 
continuum background components; $n_l$ for the $B$-background 
component is fixed to an estimate of the contribution from MC.
The parameters of the
signal and $B$-background PDFs are determined from MC and held fixed
in the final fit. The continuum background parameters for
$\mathcal{F}$ are fixed from the data with $0.1<|\Delta E|<0.3 \gev$.
The continuum background parameters for \DE\ and \mes are left free in the final fit.
Each \mes\ PDF is a Gaussian distribution for signal, and an ARGUS threshold
function~\cite{Argus} for continuum background.  Each \DE\ PDF is a
sum of two Gaussian distributions with equal means for the signal and
a first-degree polynomial for the continuum background.  The signal
and background $\cal{F}$ PDFs are Gaussian distributions with
different widths above and below the mean.  For the $B$-background
parameterizations, signal or continuum shapes are used depending on
the nature of the background.
The projections of the Region I data in $m_{ES}$, $\Delta E$ and $\mathcal{F}$ 
and the results of the fit are shown in Fig.~\ref{fig:mes_dE_fish},
demonstrating a clear signal.
The data in the plots pass a selection on the 
per-event signal-to-background likelihood ratio
formed from the other two fit variables, which has been 
optimized to give the greatest significance to the signal.

The signal yields for the regions of the Dalitz plot are
shown in Table \ref{tab:regions}. The systematic uncertainty
arises from the PDF parameters and from $B$-background subtraction.
  We find that all yields have a significance greater than five standard
deviations, where
{\footnotesize $\sqrt{ 2 \ln(\mathcal{L}_{max}/\mathcal{L}_{max (n_{signal}=0)} )}$}
is used as an estimator of the significance.

We calculate the branching fractions from the measured yields, taking 
into account the overlapping nature of the resonances, using 
${\cal B} = \mbox{M}^{-1} \ \mbox{Y}/ \nbb$ where $\mbox{Y}$ is 
a vector of the yields in each Dalitz region and $\cal{B}$ is a vector of 
branching fractions. $\mbox{M}$ is the efficiency matrix where
$\mbox{M}_{ij}$ is the probability that an event arising from 
the contribution dominating region $i$ will be found in region $j$.
The elements of $\mbox{M}$ are estimated using MC including small corrections for
differences in tracking and in particle identification efficiencies between MC and data, and
differences between MC and our resonance model.

\input models

In our resonance model we assume one dominant contribution per region.
The contributions for the chosen model are given in Table~\ref{tab:model}.
For Regions II and VI where the main contributions are not known a priori,
we take the dominant contributions to be   $K^{*0}_0(1430) \pi^+$ and 
$f_2(1270) K^+$ respectively.
However, we quote branching fractions for 
$B^+ \ra $``higher $K^{*0}$''$\pi^+$ where ``higher $K^{*0}$'' means any 
combination of $K^{*0}_0(1430), K^{*0}_2(1430)$ and $K^{*0}_1(1680)$ and
$B^+\ra$``higher $f$''$ K^+$, where ``higher $f$'' means any combination 
of $f_2(1270)$, $f_0(1370)$ and $f_2(1430)$. 
Non-relativistic Breit--Wigner line shapes are used for
all channels except for the broad $\rho (770)$ resonance, where we use a
relativistic Breit--Wigner line shape with Blatt--Weisskopf damping
\cite{BlattW}.  

We evaluate systematic uncertainties on the branching fractions
taking into consideration uncertainties on resonance parameters and 
alternative line shapes given in Table~\ref{tab:model}.  
We also include the effect on all branching fractions that in 
Regions II and VI, the dominant contribution could be any 
combination of a number of resonances.
Also included in the systematic uncertainties is the possibility that the 
yield measured in Region VII 
is from a resonant component and does not extend into the other regions.
Uncertainties on the branching fractions due to the interference between
the resonances are evaluated by generating many simulated  $B^+\ra 
K^+\pi^-\pi^+$ Dalitz plots with all the contributions having random 
phases and observing how the interference between the contributions affects
the measured branching fractions.
The branching fractions and uncertainties of intermediate resonances
 are given in Table~\ref{table:BFs}.

\input BF

\begin{figure}[htbp]
\resizebox{\columnwidth}{!}{
   \includegraphics{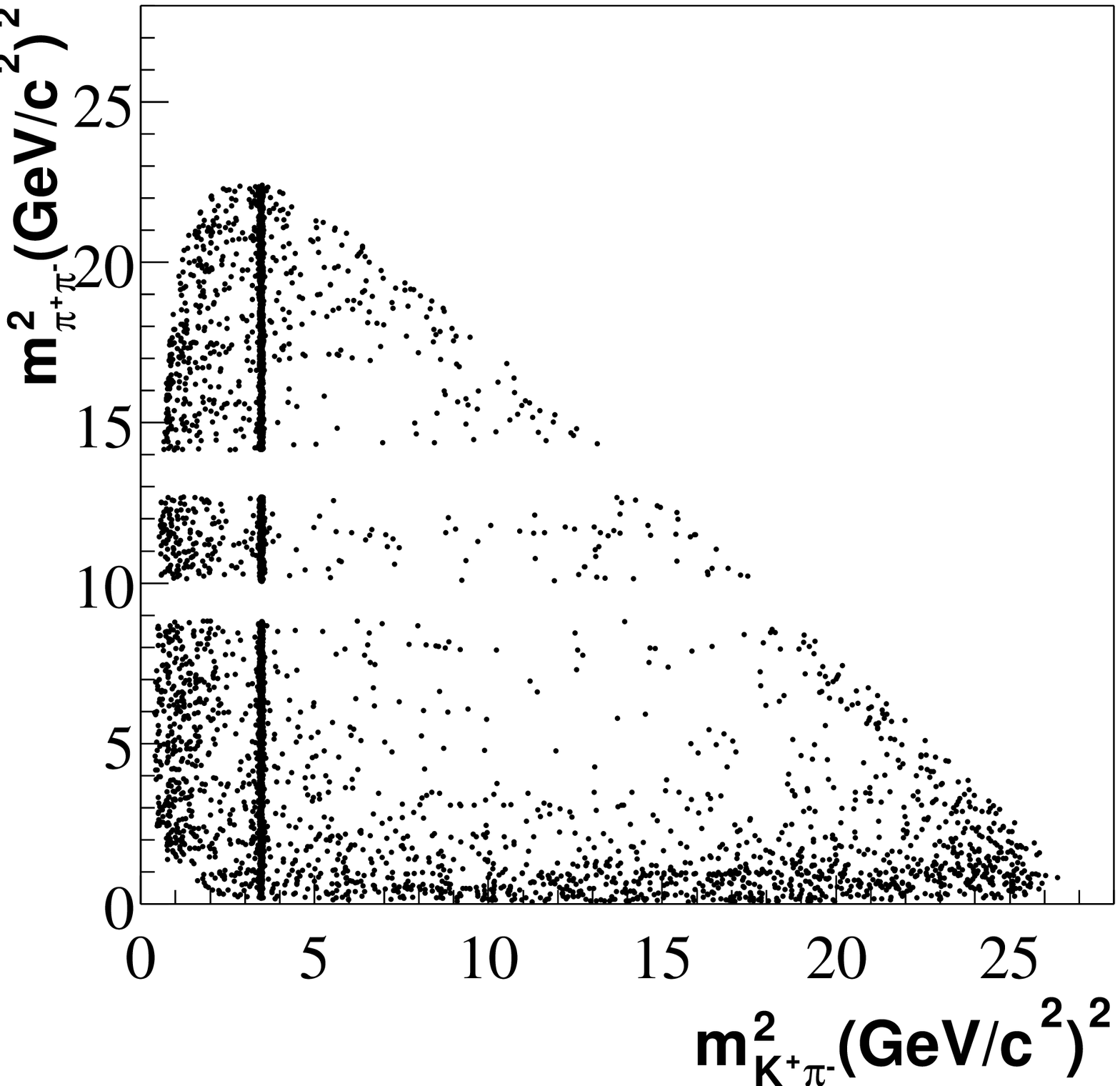}
   \includegraphics{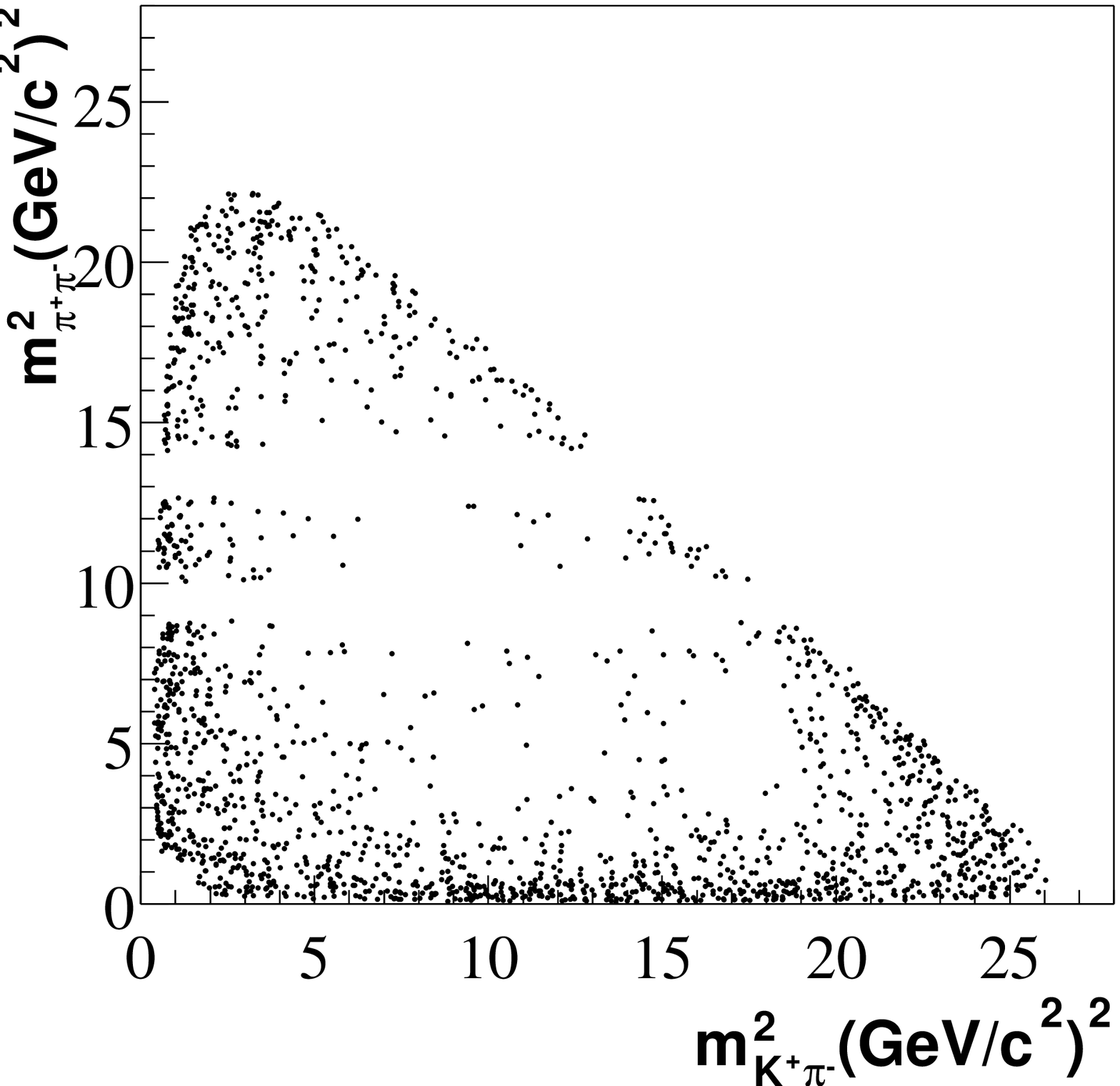}
}
\caption{Dalitz plots showing (left) the observed distribution in a signal \mes \ region (defined in the text) and (right) the distribution for continuum background from the \mes sideband. Vetoes reject events with \jpsi\ and \psitwos. 
}\label{fig:DalitzPlotData}
\end{figure}

Fig.~\ref{fig:DalitzPlotData} shows the Dalitz plot for data events
within a signal region, $5.2715 < \mes < 5.2865 \gevcc$, that have a
the per-event signal-to-background likelihood ratio
formed from the \DE\ and $\mathcal{F}$ PDFs, greater than 5.
Both signal and background
events appear in the plot. The \Dzb \pip signal is the narrow band in
the $m_{K\pi}$ spectrum.  To illustrate the expected background
distribution, events passing the same likelihood selection but having
a value of $\mes$ between $5.25 < \mes < 5.26 \gevcc$ are also shown.
The size of this sideband is chosen to contain approximately the
expected number of background events that will enter the signal-region plot.

\begin{figure}[htbp]
\resizebox{\columnwidth}{!}{
   \includegraphics{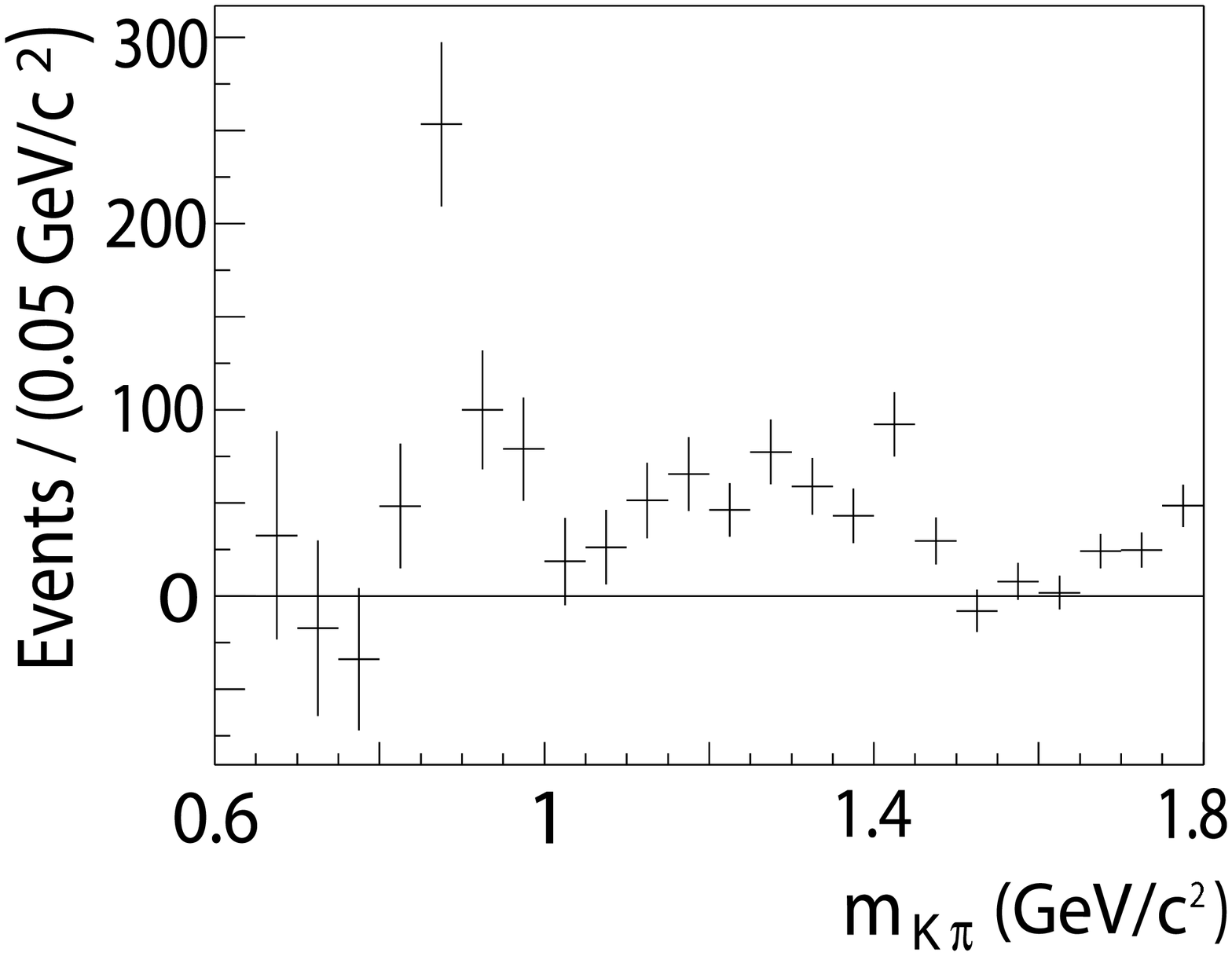}
   \includegraphics{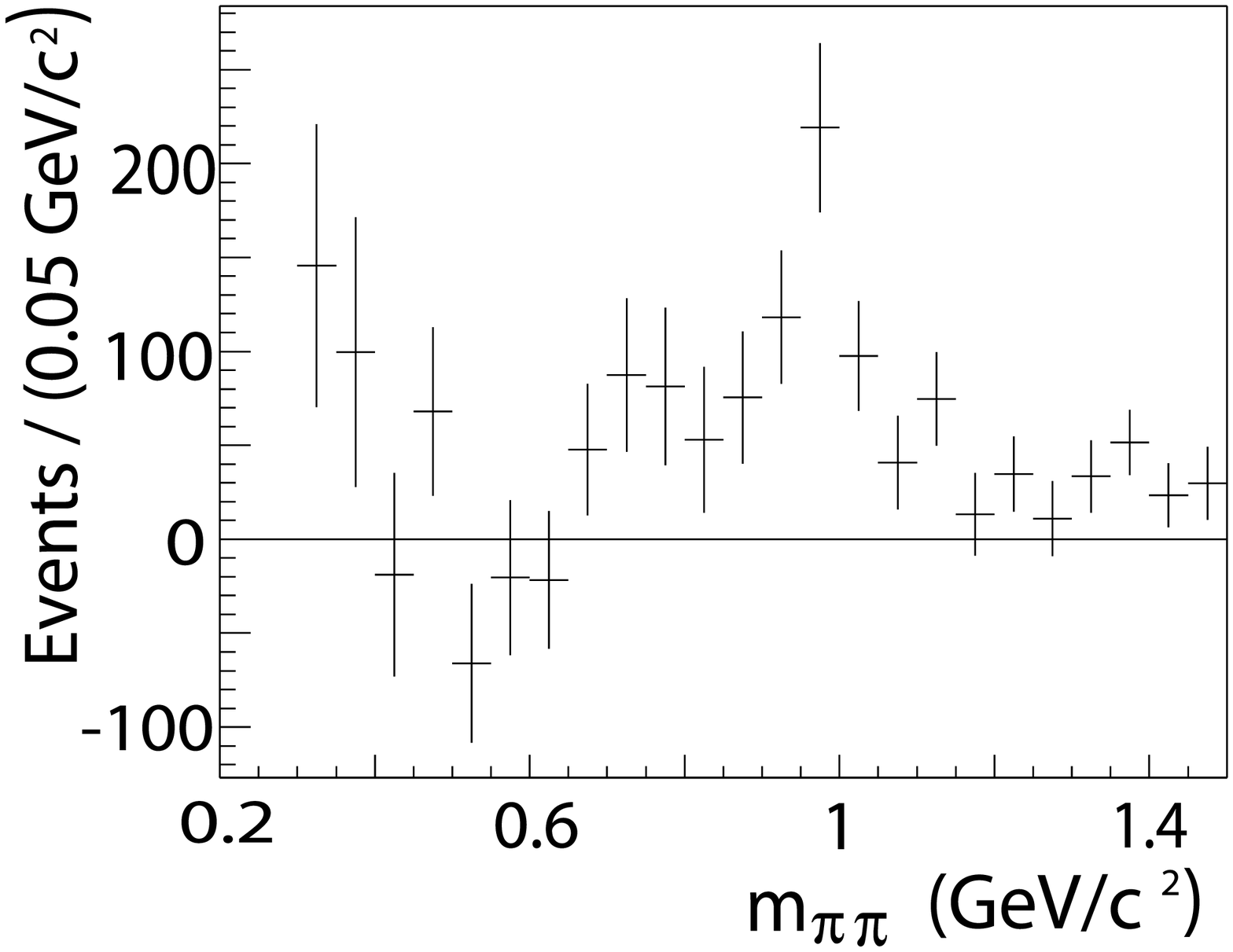}
}
\caption{
Background-subtracted and efficiency-corrected projections of the Dalitz plot in $m_{K\pi}$ and $m_{\pi\pi}$.
  }
\label{fig:ProjPlotsLongRange}  
\end{figure}

Fig.~\ref{fig:ProjPlotsLongRange} shows background-subtracted, 
efficiency--corrected projections of the two-body invariant mass spectra,
$m_{K\pi}$  and $m_{\pi\pi}$, from $0.6\gevcc$ to $1.8\gevcc$ and
from $0.2\gevcc$ to $1.5\gevcc$ respectively. 
The signal events and background distributions are obtained by the same 
method as for Fig.~\ref{fig:DalitzPlotData} and again
the $\Dzb$, $J/\Psi$ and $\Psi (2S)$ vetoes are applied.  
Peaks at the $K^{*0}(892)$ and $f_0(980)$ masses
are clearly visible.

\begin{figure}[htbp]
\resizebox{\columnwidth}{!}{
\begin{tabular}{cc}
    \includegraphics{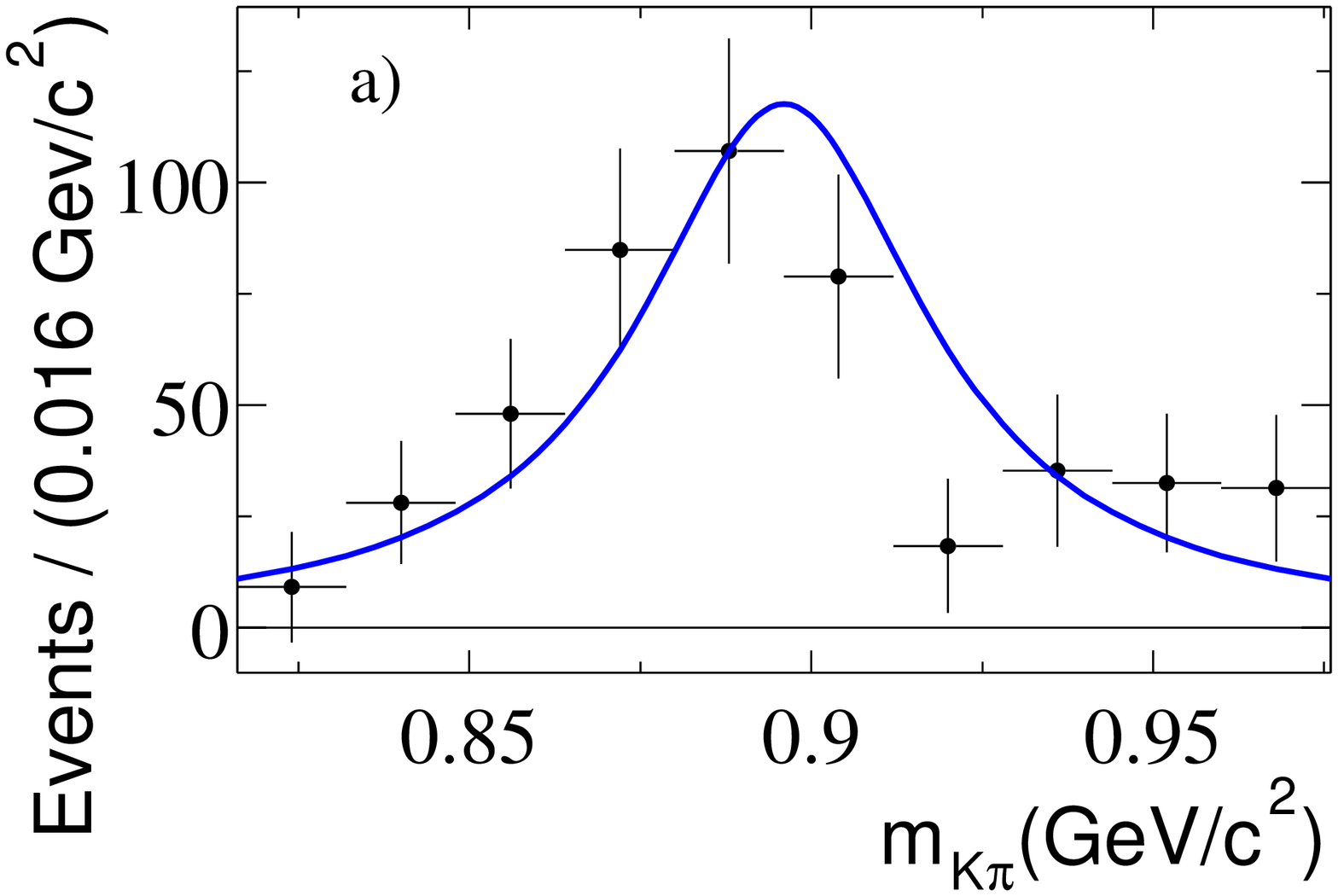} &
    \includegraphics{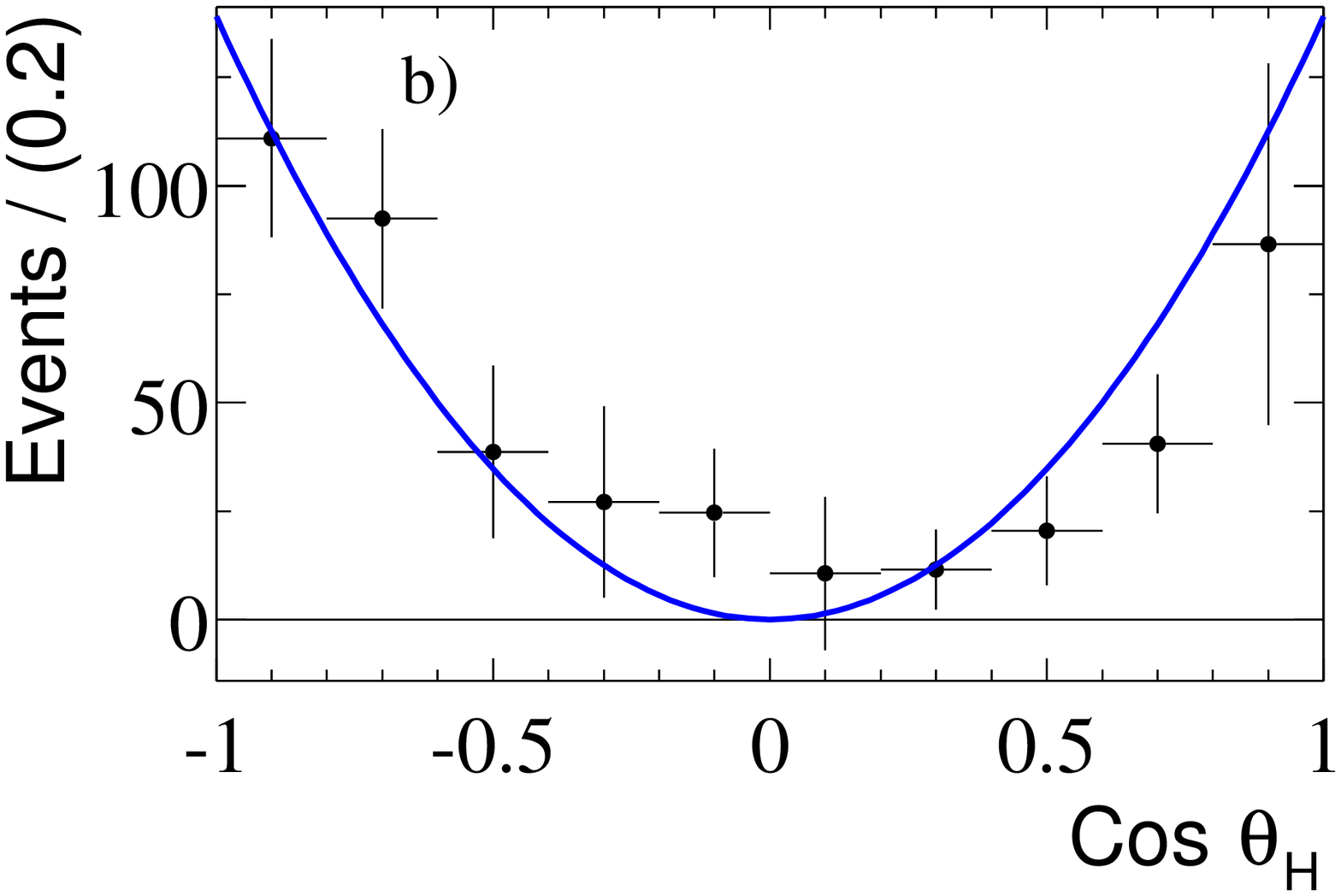} \\
    \includegraphics{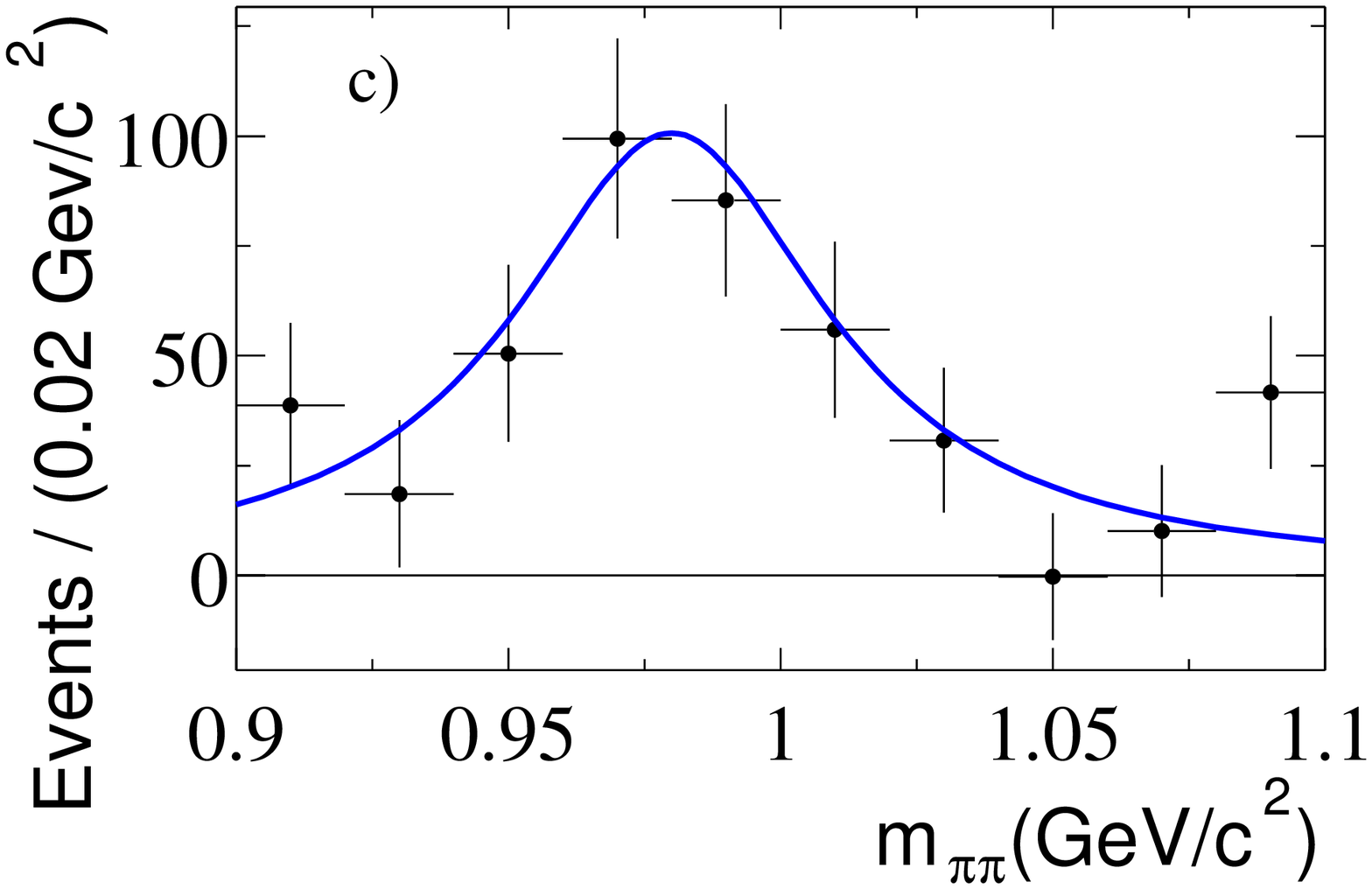} &
    \includegraphics{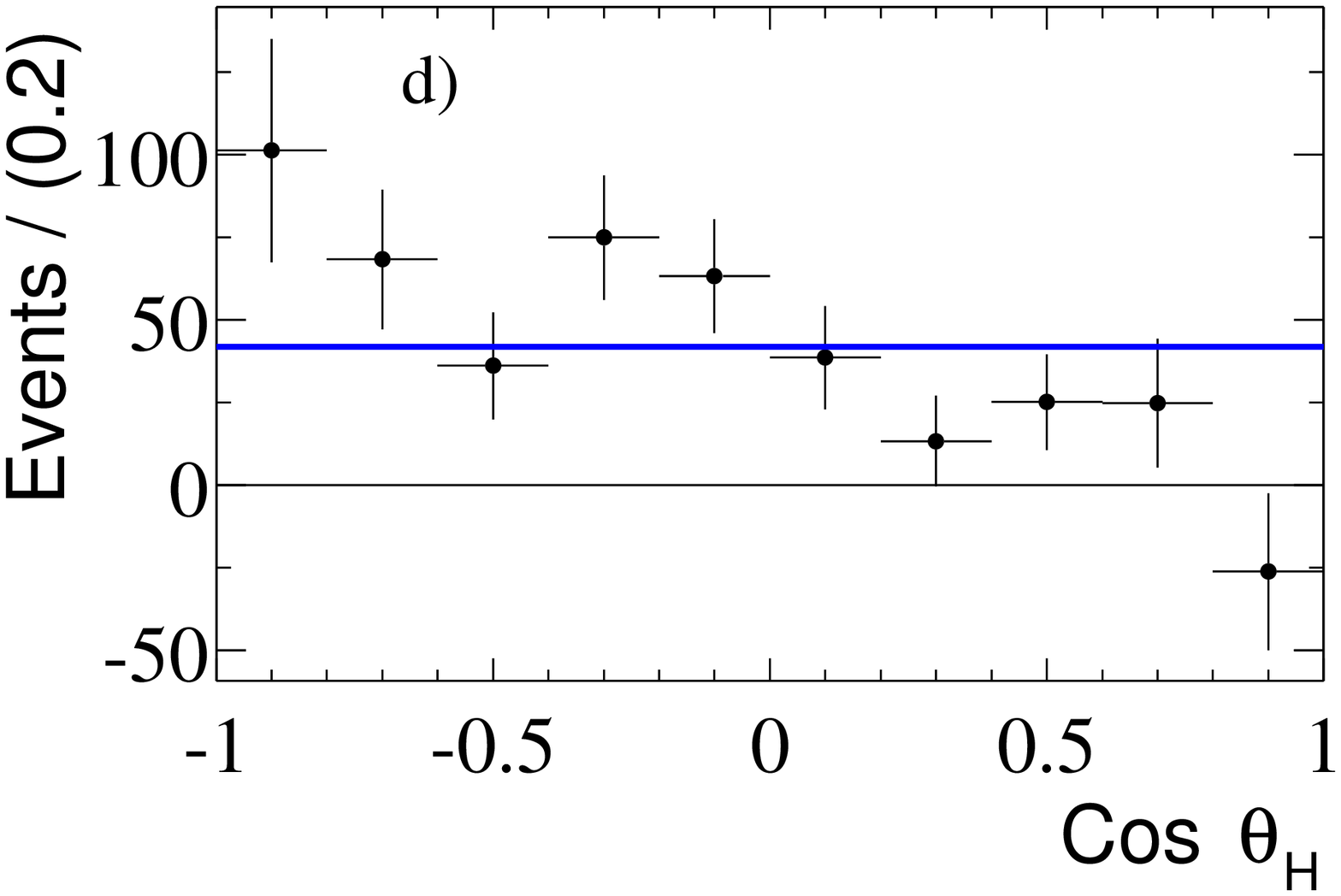} \\
    \includegraphics{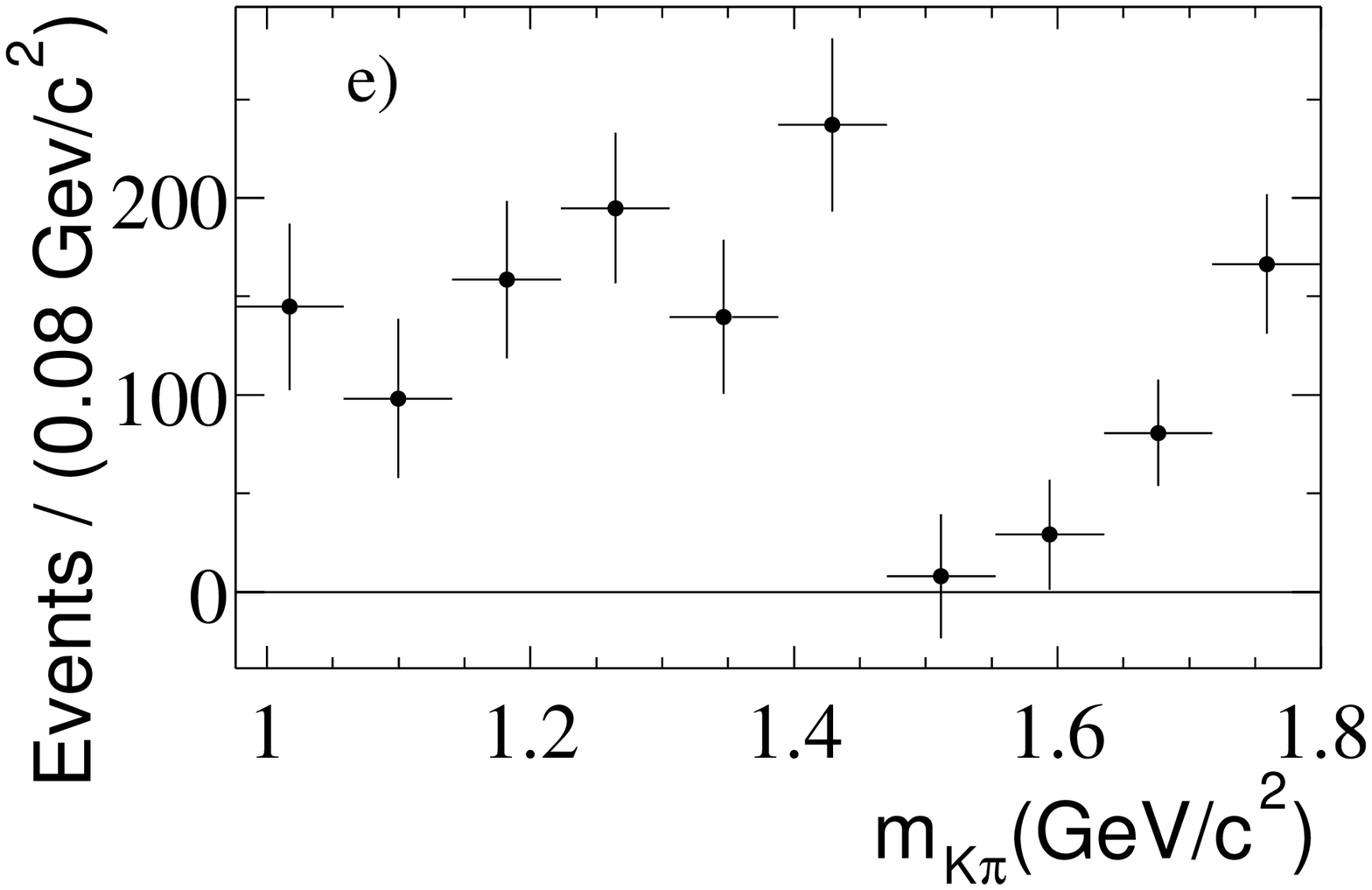} &
    \includegraphics{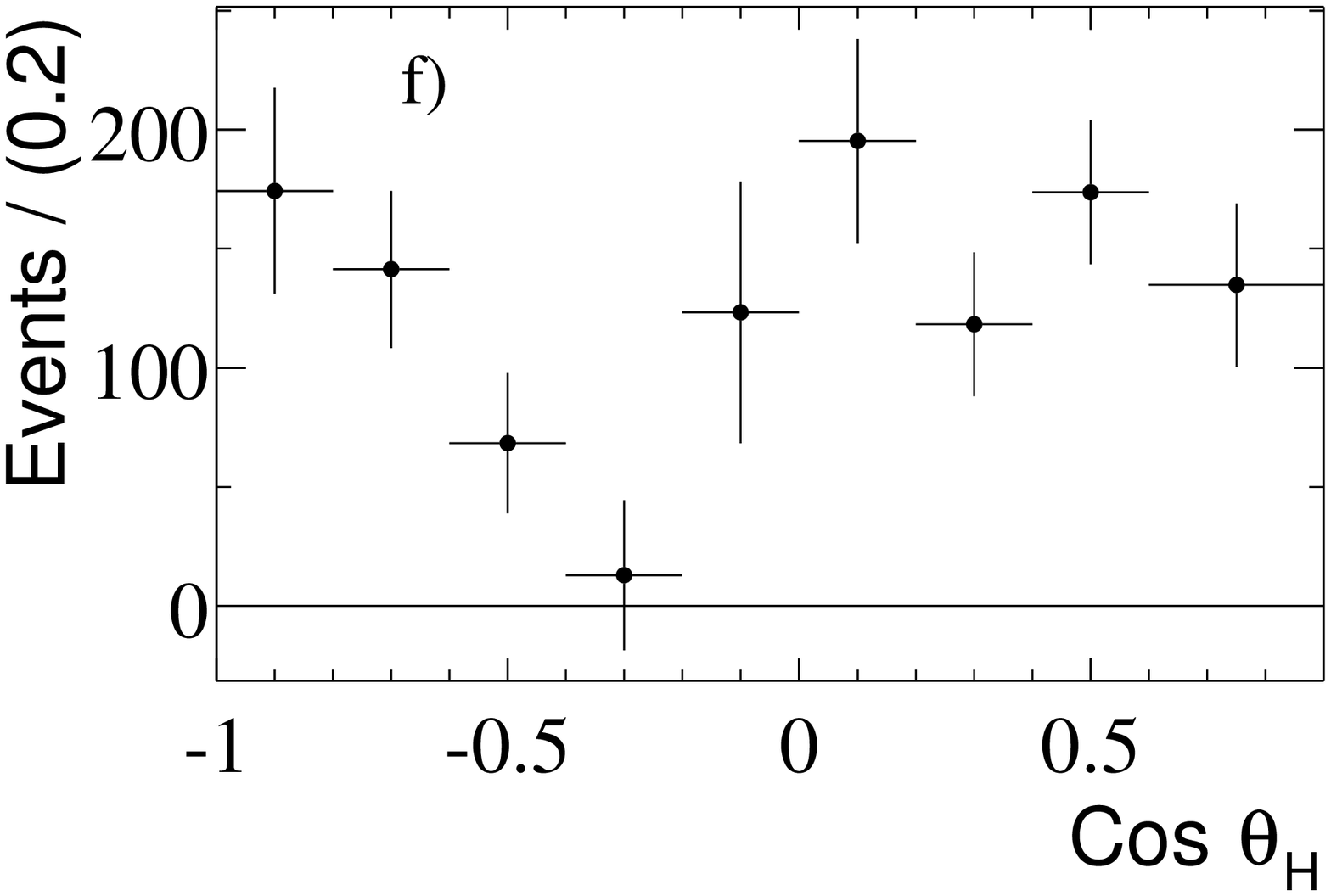} \\
\end{tabular}
}
\caption{
  Projection plots of the two-body invariant mass and $\cos\theta_H$ 
  for, from top to bottom, Regions I, V and II. 
  }
\label{fig:ProjPlotsHelRes}  
\end{figure}

The invariant mass $m_{K\pi}$ or $m_{\pi\pi}$, 
and helicity angle between the resonance decay and flight directions,
$\theta_H$, are not used in the likelihood fit. 
However, to illustrate our findings, we show, in 
Fig.~\ref{fig:ProjPlotsHelRes}, resonant mass and $\cos\theta_H$ projections
for Regions I, IV and II after background subtraction and 
efficiency corrections.
Figs.~\ref{fig:ProjPlotsHelRes}(a--d) have been overlaid with the 
distribution of the expected dominant resonance: Breit--Wigner line 
shapes for the mass distributions in
Figs.~\ref{fig:ProjPlotsHelRes}(a) and ~\ref{fig:ProjPlotsHelRes}(c), 
$\cos^2\theta_H$ for the $K^{*0}(892)$ $\cos\theta_H$ 
distribution in Fig.~\ref{fig:ProjPlotsHelRes}(b), 
and a uniform distribution for the scalar $f_0(980)$ 
$\cos\theta_H$ distribution in Fig.~\ref{fig:ProjPlotsHelRes}(d).
There is good agreement between the overlaid and observed distributions 
indicating that the expected resonances are indeed dominant in these regions.
The $f_0(980)$ $\cos\theta_H$ distribution suggests a linear dependence 
that is most likely due to interference with the vector $\rho(770)$, which is 
taken into account in our interference systematic uncertainty.

We can see in Fig~\ref{fig:ProjPlotsLongRange}, there is a large signal 
in the region $1.1<m_{K\pi}<1.4\gevcc$ (Region II). 
This is shown in more detail in Fig.~\ref{fig:ProjPlotsHelRes}(e), resonant
mass, and  Fig.~\ref{fig:ProjPlotsHelRes}(f), the  $\cos\theta_H$ projections
for Region II.
The complex behavior of this signal is 
similar to that observed by LASS~\cite{LASS} and precludes
an interpretation as a single resonance.

In conclusion, we have made branching fraction measurements, 
summarized in Table~\ref{table:BFs}, for a number of charm and 
charmless $B$ decay channels with the final state $K^+\pi^-\pi^+$.
This analysis has taken into account the uncertainty in the knowledge 
of the nature  and parameterization of the intermediate resonances on all 
the branching fractions assuming a non-resonant contribution with kinematics
defined by phase space. 
The results also take account of the unknown levels of interference between 
the different contributions.
The \BpmDzpi\ and \BpmChiczK\ results agree with previous measurements~\cite{D0br, chickbr}.
The \BpmKstarpi\ ~\cite{kstbr} and \BpmfzK\ branching fractions are consistent with, 
and more precise than, previous measurements~\cite{BelleNew}. 
The \BpmKstarpi\ result is significantly higher than predicted by many factorization
models~\cite{noel}.
The observation of the decay \BpmfzK \ provides hints about the nature of the $f_0(980)$~\cite{Minkowski:2003ja}.
A large signal is seen for $B^+ \ra $``higher $K^{*0}$''$\pi^+$ where ``higher $K^{*0}$'' means any 
combination of $K^{*0}_0(1430), K^{*0}_2(1430)$ and $K^{*0}_1(1680)$.

We also give 90\% confidence-level upper limits for the branching fractions of the
following channels:   
$\calB(\BpmRhoK) < 6.2\times 10^{-6}$,
$\calB(B^+ \ra K^+\pi^-\pi^+$ non-resonant$) < 17\times 10^{-6}$,
$\calB(B^+\ra$``higher $f$''$ K^+) < 12 \times 10^{-6}$.
The tight limit on the non-resonant component means that its
$\gamma$-dependent interference with the $\chi_{c0}K$ final state will be
very hard to measure.

\input pubboard/acknow_PRL.tex

\end{document}

%% file: symbols.tex
\newcommand{\onreslumi}  {\mbox{56.4 \invfb}}

\newcommand{\bbpairs}    {\mbox{61.6 million}}
\newcommand{\nbb}    {\mbox{$N_{\BB}$}}
\newcommand{\fz}           {\mbox{$f_0(980)$}}

\newcommand{\ppK}          {\mbox{$K^+ \pi^- \pi^+$}}
\newcommand{\Kpp}          {\mbox{$K^{\pm} \pi^{\mp} \pi^{\pm}$}}

\newcommand{\Kstarpi}      {\mbox{$\Kstarz(892) \pi^+$}}
\newcommand{\BpmKstarpi}   {\mbox{$B^+ \to \Kstarpi$}}

\newcommand{\RhoK}         {\mbox{$\rho^0(770) K^+$}}
\newcommand{\BpmRhoK}      {\mbox{$B^+ \to \RhoK$}}
\newcommand{\fzK}          {\mbox{$\fz K^+$}}
\newcommand{\BpmfzK}       {\mbox{$B^+ \to \fzK$}}

\newcommand{\ChiczK}       {\mbox{$\chi_{c0} K^+$}}
\newcommand{\BpmChiczK}    {\mbox{$B^+ \to \ChiczK$}}
\newcommand{\Dzpi}         {\mbox{$\Dzb \pi^+$}}
\newcommand{\BpmDzpi}      {\mbox{$B^+ \to \Dzpi$}}

\newcommand{\BrDzpi}{184.6 \pm 3.2 \pm 9.7}

\newcommand{\BrfzK} {9.2\pm1.2^{+2.1}_{-2.6}}

\newcommand{\BrfzKVal}      {(\BrfzK)\times 10^{-6}}

\newcommand{\BrDzpiVal}     {(\BrDzpi)\times 10^{-6}}

\newcommand{\DE}{\ensuremath{\Delta E}}

\newcommand{\calB}{\mbox{${\cal B}$}}

\newcommand{\pvec}{{\bf p}}

\def\Y#1S{{\Upsilon\rm(#1S)}}

\def\ra{\rightarrow}

\def\beq{\begin{equation}}
\def\eeq{\end{equation}}

%% file: pubboard/authors_jul2003.tex
%
\author{B.~Aubert}
\author{R.~Barate}
\author{D.~Boutigny}
\author{J.-M.~Gaillard}
\author{A.~Hicheur}
\author{Y.~Karyotakis}
\author{J.~P.~Lees}
\author{P.~Robbe}
\author{V.~Tisserand}
\author{A.~Zghiche}
\affiliation{Laboratoire de Physique des Particules, F-74941 Annecy-le-Vieux, France }
% ...
\collaboration{The \babar\ Collaboration}
\noaffiliation

%% file: dalitzTable.tex
\begin{table}[htb]
  \caption{Regions of the \ppK\ Dalitz plot and signal yields obtained (the first error is statistical and the second is systematic). The \Dzb\ band, $1.8<m_{K\pi}<1.9 \gevcc$, is excluded from all regions except region III, and the $\chi_{c0}$ band, $3.355<m_{\pi\pi}<3.475 \gevcc$, is excluded from all regions except region VIII.}
   \begin{center}
    \resizebox{\columnwidth}{!}{
    \begin{tabular}{lcl}
      \hline
      Region & Selection Criteria \ (\gevcc) & Signal Yield \\
       \hline
      I      & $0.816<m_{K\pi}<0.976$ , $m_{\pi\pi}>1.5$         & $161 \pm 18 \pm 4$        \\
      II     & $0.976<m_{K\pi}<1.8$   , $m_{\pi\pi}>1.5$         & $405 \pm 28 \pm 13$ \\
      III    & $1.835<m_{K\pi}<1.895$                            &$3755 \pm 66 \pm 11$       \\
      IV     &                          $0.6<m_{\pi\pi}<0.9$     &  $66 \pm 15 ^{+3}_{-7}$   \\
      V      &                          $0.9<m_{\pi\pi}<1.1$     & $179 \pm 19 \pm 5$        \\
      VI     &                          $1.1<m_{\pi\pi}<1.5$     & $126 \pm 19 \pm 5$        \\
      VII    & $m_{K \pi} >  1.9$     , $m_{\pi\pi}>1.5$         & $133 \pm 23 ^{+9}_{-22}$  \\ 
      VIII   & $m_{K \pi} >  1.9$     , $3.37<m_{\pi\pi}<3.46$   &  $26 \pm  6 \pm 1$        \\
      \hline
    \end{tabular}
    }
    \label{tab:regions}
   \end{center}
\end{table}

%% file: models.tex
 \begin{table}
\caption{A summary of the model used to calculate branching fractions. Alternative lineshapes are given in parentheses. The masses and widths are taken from the Review of Particle Physics~\cite{PDG}. 
}
\begin{tabular} {lccc}
\hline
Resonance & Lineshape & Mass  & Width \\ 
           &           &    (\mevcc) &  (\mevcc) \\
\hline
$K^{*0}(892)$    & BW & $896.10\pm 0.27$ & $50.7\pm0.6$ \\
$K^{*0}_0(1430)$ & BW (LASS\cite{LASS}) & $1412\pm6$ & $294\pm23$ \\
$D^0$            & BW &  $1864.5\pm0.5$ & 0  \\
$\rho^0(770)$    & Blatt-Weisskopf\cite{BlattW} & $769.0 \pm 0.9$ & $150.9\pm1.7$ \\
$f_0(980)$       & BW (Flatt\'e\cite{FLATTE}) & $980 \pm 10$ & $70 \pm 30$  \\
$f_2(1270)$      & BW & $1275\pm12$ & $185\pm30$  \\
$\chi_{c0}$      & BW & $3415.1\pm0.8$  & $16.2\pm3.2$  \\
non-resonant     & flat & all masses & - \\         
\hline 
\end{tabular}
\label{tab:model}
\end{table}

%% file: BF.tex
\begin{table}
\caption{The measured branching fractions and uncertainties. The first uncertainty is statistical, the second includes the systematic uncertainties from the yields and the efficiencies, the third is the model uncertainty, and the fourth is the uncertainty due to interference.}
\addtolength{\extrarowheight}{0.1cm}
\begin{tabular}{ll}

\hline
Channel & BF  $\times 10^{-6}$  \\ \hline 

$K^{*0}(892)\pi^+$ &  $15.5 \pm 1.8 \pm 1.1 ^{+0.6}_{-3.8} \pm 0.9$  \\ 
``higher $K^{*0}$''$\pi^+$, $K^{*0}\ra K^+\pi^-$ & $25.1 \pm 2.0 \pm 2.9 ^{+9.4}_{-0.5} \pm 4.9$  \\ 
\Dzb $\pi^+$, $\Dzb \to K^+ \pi^-$ & $184.6 \pm 3.2 \pm 9.7$   \\ 
$\rho^0(770)K^+$  & $3.9 \pm 1.2 ^{+0.3 +0.3}_{-0.6 -3.2} \pm 1.2$  \\ 
$f_0(980)K^+$, $f_0\ra\pi^+\pi^-$ & $9.2 \pm 1.2 \pm 0.6 ^{+1.2}_{-1.9} \pm 1.6$  \\ 
``higher $f$''$K^+$, $f\ra\pi^+\pi^-$ & $3.2 \pm 1.2 \pm 0.5 ^{+5.8}_{-2.4} \pm 1.5$  \\ 
Non-resonant  & $5.2 \pm 1.9 ^{+0.8+3.3}_{-1.8-7.5} \pm 6.4$  \\ 
$\chi_{c0} K^+$, $\chi_{c0}\to\pi^+\pi^-$ & $1.5 \pm 0.4 \pm 0.1$     \\ \hline

\end{tabular}   
\label{table:BFs}
\end{table}

%% file: pubboard/acknow_PRL.tex
We are grateful for the excellent luminosity and machine conditions
provided by our \pep2\ colleagues, 
and for the substantial dedicated effort from
the computing organizations that support \babar.
The collaborating institutions wish to thank 
SLAC for its support and kind hospitality. 
This work is supported by
DOE
and NSF (USA),
NSERC (Canada),
IHEP (China),
CEA and
CNRS-IN2P3
(France),
BMBF and DFG
(Germany),
INFN (Italy),
FOM (The Netherlands),
NFR (Norway),
MIST (Russia), and
PPARC (United Kingdom). 
Individuals have received support from CONACyT (Mexico), A.~P.~Sloan Foundation, 
Research Corporation,
and Alexander von Humboldt Foundation.

%% file: paper_eprint.bbl
\begin{thebibliography}{99}
\bibitem{kstbr}
The comparison takes into account that the secondary branching fraction
of $K^{*0}\ra K^+\pi^-$ is $2/3$.
\end{thebibliography}
